\documentstyle[preprint,aps]{revtex}
\draft
\begin{document}
\title{First and Second Sound Modes of a Bose-Einstein Condensate in a Harmonic
Trap}
\author{V.B. Shenoy and Tin-Lun Ho}
\address{Department of Physics,  The Ohio State University, Columbus, Ohio
43210}
\maketitle

\begin{abstract}
We have calculated the first and second sound modes of a dilute 
interacting Bose gas in a spherical trap for 
temperatures ($0.6<T/T_{c}<1.2$) and for systems with $10^4$ to $10^8$
particles.
The second sound modes (which exist only below $T_{c}$) generally 
have a stronger temperature dependence than the first sound modes. The 
puzzling 
temperature variations of the sound modes near $T_{c}$ recently 
observed at JILA in systems with $10^3$ particles match surprisingly well 
with those of the first and  second sound modes of much larger systems. 
\end{abstract}


Since the discovery of Bose-Einstein condensation in atomic gases of alkali
atoms\cite{Rb}, there has been great interest in the broken gauge symmetry 
({\em i.e.,} the ``phase") of the condensate. In the case of $^4$He, its 
``phase" dynamics leads to the existence of second sound, which is essentially
the out of phase pressure and temperature oscillations. In a series of sound 
experiments, Jin et.al at JILA\cite{Jin} have observed a number of 
``puzzling" behaviors in the temperature dependence and the dissipation 
of the sound modes above $0.5T_{c}$. There are no explanations for these
behaviors so far.  Jin et.al. have speculated that the observed ``$m=0$" mode 
could be the ``second sound". If
this were true, it would be a demonstration of the broken gauge symmetry of 
the system. However, 
in the absence of a detailed calculation consistent with experiments, 
the identification of the second sound mode would be difficult.

To help identify the nature of the sound modes, we have solved the linearized 
two-fluid hydrodynamic equations of an interacting dilute Bose gas in a
spherical harmonic trap. 
We have in mind systems that are sufficiently large so that the hydrodynamic
approach is accurate\cite{Pethick}. It should be noted 
that the recent experiments at JILA\cite{Jin}  were performed on small
systems with a few thousand atoms. While the hydrodynamic modes of a
large system may be different from the sound modes of a small one, the
study of the former is important in its own right. After all, the 
number of atoms in the Bose condensate has
increased from $10^3$ to $10^6$ within six months after the initial
discovery\cite{Rb}. It would not be surprising if Bose condensates with $10^9$
atoms were produced in the near future. 
On the other hand, the hydrodynamic modes of large 
systems {\em are} relevant for the sound modes of small ones, as 
the former must evolve smoothly into the latter as the number of atoms
is decreased continuously. This suggests the
possibility of identifying the nature of sound modes of a small system by
studying their hydrodynamic counterparts in a large one. 
Indeed, comparing our results (for systems with $10^4$ to $10^8$ atoms)
with the JILA observations\cite{Jin}, we find that
the temperature variations of the observed sound modes
 show up in the analogous modes of the larger systems in a spherical trap.
In particular, the ``mysterious" behaviors  of the JILA $(m=0)$ 
and $(m=2)$ modes\cite{Jin} in the range
$0.5<T/T_{co}<0.8$ match closely with the behaviors of the second sound modes
of the larger systems in same temperature range, while the frequency 
and temperature dependence of the observed $(m=0)$ mode above $T_{c}$ 
are identical to those of the first sound mode in the same  temperature range. 
(The temperatures $T_{co}$ and $T_c$ are the transition temperature for the 
ideal Bose and the dilute interacting Bose gas respectively.)\cite{Griffin}

Our choice of spherical symmetry is to keep the calculations manageable. 
Moreover, as a first step, we shall ignore dissipation. While it
is entirely feasible within our scheme to include dissipative effects, 
we feel that it is important (as in bulk $^{4}$He) to first understand 
dissipationless hydrodynamics, so that one can clearly identify the 
dissipative effects later in a complete solution. 

{\bf Linearized Hydrodynamics :} 
We begin with the two-fluid hydrodynamic equations of Bosons with mass
$M$ in an external potential $\phi({\bf r})$\cite{HS}, 
$M\dot{n} = -{\bf \nabla}\cdot{\bf g}$,
$\dot{g}_{i} = -n\nabla_{i}\phi  - \nabla_{j}\Pi_{ij}$,
$\dot{s} = -{\bf \nabla}\cdot(s{\bf v}_n)$, and 
$\dot{\bf v}_{s} = -\frac{1}{M}{\bf \nabla}(\mu + \phi + M{\bf v}_n\cdot{\bf
v}_s)$,  
where $n$, ${\bf g}$, $\Pi_{ij}$, $s$, $\mu$ 
are the number density, the momentum density, the stress tensor, the 
entropy density, and the 
chemical potential respectively. Here, ${\bf v}_{n}$ and 
${\bf v}_s \equiv (\hbar/M){\bf \nabla}\theta$ are the normal fluid 
and superfluid velocities respectively, 
where $\theta$ is the phase of condensate. 
For a spherical harmonic trap with frequency
$\omega_{T}$, $\phi(r)  = \frac{1}{2}M\omega_{T}^2 r^2$.
In the presence of a condensate, $n$, ${\bf g}$, and $\Pi_{ij}$ 
are of the form $n = n_{s}+ n_{n}$, 
${\bf g} = M(n_n{\bf v}_n + n_s{\bf v}_s)$, 
and $\Pi_{ij} = P\delta_{ij} + M(n_n v_{ni}v_{nj} + n_s v_{si}v_{sj})$, where
$n_{s}$ and $n_{n}$ are the superfluid and normal fluid number densities, 
and $P$ is the pressure. Denoting
the equilibrium quantities by the subscript ``$_{o}$", we have 
${\bf v}_{no} = {\bf v}_{so} = 0$,
${\bf \nabla}P_{o} + n_{o}{\bf \nabla}\phi = 0$, and
${\bf \nabla}(\mu_{o}+\phi) = 0$.
Using the Gibbs-Duhem relation 
$d \mu = -\sigma d T + d P/n$, where $\sigma \equiv s/n$ is the entropy 
per particle, these equilibrium conditions imply 
${\bf \nabla}T_{o} = 0$, and hence 
\begin{equation}
{\bf \nabla} n_{o} = \left(\frac{\partial n}{\partial P}\right)_{T_{o}}
{\bf \nabla}P_{o} =  - n_{o} \left(\frac{\partial n}{\partial P}
\right)_{T_{o}}{\bf \nabla}\phi,  \,\,\,\,\,\,\,\,\,\,
{\bf \nabla} \sigma_{o} = \left( \frac{\partial \sigma_{o}}{\partial
P}\right)_{T_{o}} {\bf \nabla}P_{o} = -\frac{1}{n_{o}}
\left( \frac{\partial n}{\partial T}\right)_{P_{o}}{\bf \nabla}\phi , 
\label{edensity} \end{equation}
where we have made use of the Maxwell relation $(\partial \sigma/\partial
P)_{T} = n^{-2} (\partial n/\partial T)_{P}$. 

Denoting the deviation of any 
quantity $x$ from its equilibrium value $x_{o}$ as $\delta x \equiv x -
x_{o}$, the hydrodynamic equations can be linearized about 
the equilibrium solution and written as 
${\bf I :}$ $\delta \dot{n}$$=-{\bf \nabla}\cdot( n_{o}{\bf v}_{s}$
$+n_{no}{\bf w})$, 
${\bf II :}$$n_{o}\dot{\bf v}_{s}$
$+n_{no}\dot{\bf w}= -(\delta n {\bf \nabla}\phi$
$+{\bf \nabla}\delta P)/M$, 
${\bf III :}$ 
$\dot{\bf v}_{s}=\frac{1}{M}{\bf \nabla}(\sigma_{o}\delta T$
$-\frac{\delta P}{n_{o}})$, 
${\bf IV :}$
$\delta \dot{s}= -{\bf \nabla}\cdot \left(s_{o}{\bf v}_{n}\right)$, 
where ${\bf w}={\bf v}_{n}-{\bf v}_{s}$. Using eq.(\ref{edensity}), 
{\bf I} and {\bf II} imply that 
\begin{equation}
M \delta \ddot{n} = 
{\bf \nabla}\cdot \left[ n_{o} {\bf \nabla}
\left( \frac{\delta P}{n_{o}}\right) - 
\delta T n_{o}{\bf \nabla}\sigma_{o} \right] \equiv A . 
\label{A} \end{equation}
Again using eq.(\ref{edensity}), {\bf II} and {\bf III}, we have 
$\dot{\bf w}= -\frac{ n_{o}\sigma_{o}}{M n_{no}}{\bf \nabla}\delta T$. 
By noting that $\dot{s} = \sigma_{o} \dot{n}
+ n_{o}\dot{\sigma}$, it is easy to show from ${\bf IV}$ that 
$\ddot{\sigma} = -\dot{\bf v}_{n} \cdot {\bf \nabla}\sigma_{o}
-\frac{\sigma_{o}}{n_{o}} {\nabla}\cdot(n_{so}\dot{\bf w})$, and hence 
\begin{equation}
M \ddot{\sigma} = -({\bf \nabla}\sigma_{o})^{2}\delta T + 
\frac{1}{n_{o}}{\bf \nabla}\cdot \left(
\frac{n_{o}n_{so}\sigma^{2}_{o}}{n_{no}}{\bf \nabla}\delta T\right) 
+ {\bf \nabla}\sigma_{o}\cdot {\bf \nabla}\left( \frac{\delta
P}{n_{o}}\right) \equiv B  . 
\label{B} \end{equation}
Expressing all quantities in terms of $\delta T$ and $\delta P$, 
Eqs.(\ref{A}) and (\ref{B}) form a 
closed set : 
$(\frac{\partial n}{\partial P})_{_T} \delta \ddot{P} + 
(\frac{\partial n}{\partial T})_{_P} \delta \ddot{T} = A/M$, 
$(\frac{\partial \sigma}{\partial P})_{_T} \delta \ddot{P} +
(\frac{\partial \sigma}{\partial T})_{_P} \delta \ddot{T} = B/M$.  
The solutions of these two equations are the hydrodynamic modes of the system. 
To find them, we need to evaluate the thermodynamic 
quantities in these equations. 

{\bf Thermodynamics :} 
The thermodynamics of a trapped dilute Bose gas has been worked out within the 
local density approximation (LDA) by Chou, Yang and Yu (CYY)\cite{CYY}. As
pointed out by CYY, LDA is valid if 
(a) $\epsilon\equiv \hbar \omega/k_{B}T<<1$, and (b) $\lambda >>a$, where 
$\lambda\equiv\sqrt{2\pi\hbar^{2}/Mk_{B}T}$ is the thermal wavelength and 
$a$ is the s-wave scattering length. 
These conditions are satisfied over a very wide range of temperatures 
above and below $T_{c}$ for large clouds. 
As discussed in CYY, LDA describes the physics over scales 
greater than $a_{T}\equiv \sqrt{\hbar/M\omega_{T}}$, and 
{\em all structures on the scale of $a_{T}$ and smaller are shrunk to a
point}. 
Since the typical width of the interface between the condensate and 
the normal gas is less than $a_{T}$ \cite{GPS}, 
it is treated as a surface (say, at $r=r^{\ast}$) 
within this scheme. 
The density is continuous at $r^{\ast}$ but its slope is not\cite{CYY}. 
Because of this discontinuity, it is necessary to find the
boundary conditions relating the solutions inside and outside 
$r^{\ast}$. 
The identification of these boundary conditions is the key to our 
numerical approach, and will be addressed in 
{\em Numerical Methods} below. 

Further simplification can be made within the temperature range 
$a\lambda^2 n_{o}\ll 1$ (denoted as condition (c)), where 
all thermodynamic quantities have been worked out in the classic work of Lee
and Yang\cite{LY}. Condition (c) is more restrictive than (a) above. However, 
it still covers a wide range of temperatures. (For a gas of 
$^{87}$Rb with $N\sim 10^6$ in a trap with $\omega_{T} \sim 10^3$sec$^{-1}$, 
the condition 
(c) is satisfied over the range $0.6<T/T_{c}<1.2$ that we studied). 
The coefficients in 
Eqs.(\ref{A}) and (\ref{B}) can be calculated in a straightforward manner 
from ref.\cite{CYY} and \cite{LY}. 
We shall not present the details here for length reasons. 
Instead, we outline the procedure and give the final expressions : 

\noindent (i) To determine the density profile at temperature $T_{o}$ below
$T_{c}$, we first specify the chemical potential 
$\mu_{o}$.
This immediately  determines the size of the condensate droplet 
$r^{\ast}$ through the relation 
$\mu_{o} = \phi(r^{\ast}) + 2g n_{c}(T_{o})$, where $n_{c}(T) \equiv 
\lambda^{-3}g_{3/2}(1)$. 
(ii) The region $r<r^{\ast}$ consists of both the condensate and 
the normal components, with 
$n_{no}(r) = n_{c}(T_{o})$, and $n_{so}(r) = g^{-1}[ \phi(r^{\ast}) - \phi(r) ]$. 
The region $r>r^{\ast}$ consists of only the normal component, with 
$n_{o}({\bf r})$  determined self consistently from the relation 
$n({\bf r}) = \lambda^{-3}_{o} g_{3/2}(\zeta({\bf r}))$, where 
${\rm ln}\zeta({\bf r}) = \beta[\mu_{o} - \phi(r) - 2g n_{o}({\bf r})]$. 
The quantities $N$ and $\mu_{o}$ are related though the constraint 
 $N = 4\pi\int {\rm d}r r^2
[ n_{so}(r) + n_{no}(r)]$. This relation combined with the condition
$\mu_{o}=2gn_{c}(T_{c})$ gives $T_{c}$ as a function of $N$. 

\noindent (iii) When $a\lambda^2 n_{o}\ll 1$, 
the Hemholtz free energy is 
$f(n({\bf r}), T) = -k_{B}T\lambda^{-3}g_{5/2}(1)
+ (g/2)[n({\bf r})^2 + 2n({\bf r})n_{c}(T) - n_{c}(T)^2 ]$ for $r<r^{\ast}$, 
and 
$f(n({\bf r}), T) =
-k_{B}T\lambda^{-3}g_{5/2}(\zeta({\bf r})) + k_{B}T n({\bf r}) 
{\rm ln}\zeta({\bf r}) + g n({\bf r})^2$ for $r>r^{\ast}$\cite{LY}. 
From these two equations, it is straightforward to calculate all
the thermodynamic quantities needed in Eqs.(\ref{A}) and (\ref{B}): 
for example, 
$P = -f + n(\partial f/\partial n)_{T}$ and 
$\sigma = s/n = -n^{-1}(\partial f/\partial T)_{n}$, and so forth.

{\bf Numerical Method} : Since Eq.(\ref{A}) and (\ref{B}) are smooth
inside and outside $r^{\ast}$, it can be solved in each of 
these regions in a standard way by discretizing it on a grid which is made 
finer as $r\rightarrow r^{\ast}$. Because 
some coefficients of these equations are discontinuous at $r^{\ast}$, (as
explained before), the solution of these equations 
must be understood mathematically in terms of the following 
limiting process. Firstly, we note that the hydrodynamic 
equations {\bf I} to {\bf IV}
are independent of the specific form of the free energy $f(T,n)$. We can then 
imagine solving these equations for a family of free energies 
$f(T,n;\tau)$ parametrized by a variable $\tau$,  which 
changes from the true free energy 
$f^{\rm true}$ to the LDA free energy $f^{LDA}$ in a smooth
manner as $\tau$, say, varies from 0 to 1. Such a change, of course,
amounts to gradually collapsing the actual interface into a very thin 
region centered at $r^{\ast}$. During the process of collapse, 
$\delta P$ and $\delta T$ are smooth everywhere. The most rapid changes takes 
place at the boundary of the interface (which we denote as
$r^{\ast}\pm\Delta$), while $\delta P$ and $\delta T$ and their 
derivatives are smooth in a close neighborhood $(r^{\ast}-\epsilon, 
r^{\ast}+\epsilon)$ of $r^{\ast}$ $(\epsilon<<\Delta)$ 
during the collapsing process.  
This implies the following (almost unique) way
to implement the boundary condition : The interface is modeled by three points
 $r^{\ast}\pm\Delta$ and $r^{\ast}$, where $\Delta$ is now the grid spacing. 
Both $\delta P$ and $\delta T$ {\em as well as} their derivatives 
are continuous at $r^{\ast}$, 
while the values of $\delta P$ and $\delta T$ 
at $r^{\ast}+\Delta$ and $r^{\ast}-\Delta$ are determined by their 
solutions inside and outside $r^{\ast}$. Thus, if sharp changes ever occur 
in the solution, they will only occur at $r^{\ast}\pm\Delta$ and nowhere else. 

As we shall see in {\bf A} below, our numerical method 
reproduces a set of non-trivial modes which 
can also be obtained analytically. These modes exist in each angular
momentum sector and are {\em independent of the validity of LDA}. 
The fact that our calculations 
reproduce the exact results for each $\ell$ shows that 
the boundary conditions have been implemented correctly.\cite{accurate}

{\bf Main Results} : Because of spherical symmetry, $\delta P$ and $\delta T$ 
can be decomposed as 
$(\delta P({\bf r}), \delta T({\bf r}))= \sum_{\ell,n,m}
(\delta P_{\ell n}(r), \delta T_{\ell n}(r))
Y_{\ell m}(\hat{\bf r})$, where $\ell$ is the angular momentum and $n$ is the
radial quantum number. Each $(\ell, n)$ mode is $2\ell+1$ fold degenerate. 
We have performed calculations for $^{87}$Rb ($a = 58.2\AA$) for the
number of trapped atoms $N$ varying from $N=10^4$ to 
$10^8$. The general features of all these systems are identical. 
For all cases studied, we find that {\em the frequencies of all first and 
second sound are above the trap frequency
$\omega_{T}$}\cite{Griffin}, 
which is to be
expected if the system is viewed as a mechanical system with internal degrees
of freedom in a harmonic potential. 
For concreteness, we present the results for 
$N=10^{6}$ particles in a trap, with $\omega_{T}/2\pi = 200$Hz, over 
the range $0.6<T/T_{c}<1.2$ : 

{\bf (A)} First sound : These modes exist both above and below $T_{c}$. 
They are {\em in phase} pressure and temperature oscillations that
extend over the entire cloud, and $\delta P$ has one more node
than $\delta T$. 
The frequencies of these modes (denoted as 
$\omega^{(1)}_{\ell n_{1}}$) are shown in Fig.1 for $\ell = 0, 1,
2$ and $n_{1}= 0$ to 4, where $n_{1}$ counts the number of nodes of $\delta P$ 
in the radial 
direction. While $\{ \omega^{(1)}_{\ell n_{1}}\}$ 
change with temperature, their variations are
small compared to $\omega_{T}$. The eigenfunctions of 
the $(\ell=1, n_{1}=2)$ mode at $T=0.84T_{co}$ are shown in Fig.2a. 
They extend over the entire cloud -- a feature common to all 
first sound modes above and below $T_{c}$.

The $n_{1}=0$ modes are special. They are isothermal modes 
of the form
$\delta P(r, \hat{\bf r})  = n_{o}(r) r^{\ell}Y_{\ell m}(\hat{\bf r})$, 
$\delta T=0$, with 
$\omega^{(1)}_{\ell,0}= \omega_{T}\sqrt{\ell}$.
They are also ``universal" in the sense that they are 
{\em independent of interaction and statistics}.
{\em These results emerge from our numerical solutions} but can also 
be obtained 
analytically from Eqs.(\ref{A}) and (\ref{B}) . With $\delta T=0$, using
the equilibrium relations given by Eq.(\ref{edensity}), Eqs.(\ref{A}) and
(\ref{B}) can be shown to yield
${\bf \nabla}^2 (\delta P/n_{o}) =0$, and
$\partial^{2}_{t} (\delta P/n_{o}) = 
-{\bf \nabla}\phi\cdot{\bf \nabla}(\delta P/n_{o})$, which has the solution 
given above.  From the hydrdodynamic equations {\bf I} to {\bf IV}, it 
is also straightforward to show that for these isothermal modes, 
${\bf v}_{n}={\bf v}_{s}$ 
below $T_{c}$, and ${\bf \nabla}\cdot{\bf v}_{n}=0$ both above and below 
$T_{c}$. 

The $(\ell=0, n_{1}=1)$ mode is also special. Above $T_{c}$, it is a uniform
temperature oscillation, ${\bf \nabla}\delta T=0$, but with $\delta T\neq 0$.  
This mode is ``non-universal" because it depends on interactions. 
The interaction effect, 
however, is sufficiently weak so that $\omega^{(1)}_{0,1}$ is very close to 
$2\omega_{T}$ above $T_{c}$. It is also straightforward to show that 
${\bf \nabla}\cdot {\bf v}_{n}$ is constant but nonzero for this mode. 
These results can be established analytically using LDA and also 
emerge as part of our numerical solutions. 
Below $T_{c}$, $\delta T$ is no longer uniform, and 
${\bf \nabla}\cdot {\bf v}_{n}$ is not a constant\cite{GWS}. 
All the other sound modes $(\ell, n_{1}\neq 0)$
are non-isothermal. 

{\bf (B)} Second Sound : These modes only exist below $T_{c}$.
The frequencies of these modes for ($\ell = 0, 1, 2$) and 
($n_{2} = 0,1, 2$) are shown in Fig.1. 
{\em It should be stressed that the second sound frequencies do 
not merge into the first
sounds frequencies as $T\rightarrow T_{c}$}. To illustrate this clearly, we plot
$\omega^{(2)}_{\ell, n_{2}=1}$ near $T_{c}$ (for $\ell=0$ to 2) as a function 
of particle number $N$ in Fig.3. While $\omega^{(2)}_{\ell, n_{2}=1}$ changes 
with $N$, 
the first sound frequencies (not shown in Fig.3) typically vary by about 
$2\%$ of $\omega_{T}$ in the same range of $N$. 
The eigenfunctions of the $(\ell=1,n_{2}=2)$ mode are 
shown in Fig.2b. An enlarged  structure of the interface of this mode 
at $r^{\ast}$ is shown in Fig.2c.  The second sound modes have the following
common features : 

\noindent {\bf (1)} $\delta P$  and $\delta T$ are ``out of phase" inside the
condensate and become ``in phase" as they leak out into the normal region. 
The leakage reduces to zero as $T\rightarrow T_{c}$. 
The quantum number $n_{2}$ counts the number of nodes of 
$\delta P$ or $\delta T$ $inside$ the condensate.  
{\bf (2)} The wavelengths of the
oscillations shrink as $r\rightarrow r^{\ast}$. This can be understood simply 
from LDA by recalling that the second sound velocity $c_{2}$ 
of a homogenous dilute Bose gas is proportional to $\sqrt{n_{so}}$. 
The wavelength $2\pi k^{-1}$ is then 
$2\pi c_{2}/\omega \propto \sqrt{n_{so}}$. 
Since $\omega\sim\omega_{T}$ in our case, 
and $n_{so}(r)$ vanishes as $r\rightarrow r^{\ast}$, the local 
wavelength shrinks as $r\rightarrow r^{\ast}$. 
{\bf (3)} The $n_{2}=0$ modes are different from all other $n_{2}\neq 0$ modes. 
Firstly, except very clost to $T_c$, 
all $\omega^{(2)}_{\ell, n_{2}=0}$ increase as $T$ decreases, whereas 
all $\omega^{(2)}_{\ell, n_{2}\neq 0}$ have opposite behavior, (see Fig.1). 
Secondly, while
$\delta T$ and $\delta P$ are out of phase for all second sound modes 
(i.e. $n_{2}=0$ and $n_{2}\neq 0$), 
the sign of ${\bf v}_{s}\cdot{\bf v}_{n}$ for a particular mode depends
on position. In particular, ${\bf v}_{s}\cdot{\bf v}_{n}$ 
of the $n_{2}=0$ modes is actually positive (i.e. in phase) 
almost everywhere inside the condensate instead of negative\cite{Griffin}, 
whereas it can be 
positive or negative for the $n_{2}\neq 0$ modes. (The radial 
components of ${\bf v}_{s}$ and ${\bf v}_{n}$ for the $n_{2}\neq 0$ modes 
are out of phase in most regions in the condensate, so are their tangential 
components. However, the in-phase and out-of-phase regions of these two 
components do not coincide.)
This shows that unlike the second sound modes in homogenous systems, 
which are characterized by either out of phase ($\delta P$, $\delta T$), or 
out of phase (${\bf v}_{s}$, ${\bf v}_{n}$) oscillations, 
{\em the correct characterization of the second sound modes 
in the trap is the out of phase 
$\delta P$ and $\delta T$ oscillations, not the out of phase 
${\bf v}_{s}$ and ${\bf v}_{n}$ oscillations}\cite{Griffin}. 
{\bf (4)} Near the center of the cloud, the ratio
$\xi\equiv|n_{n}v_{n}/n_{s}v_{s}|$ is about 0.2 to 0.3 for the modes studied. 
This is very different from $^{4}$He, where the normal current is essentially
cancelled by the supercurrent, i.e. 
$\xi\equiv|n_{n}v_{n}/n_{s}v_{s}|\sim 1$\cite{Griffin}. That $\xi$ 
is between 
0.2 and 0.3 can be understood in terms of LDA.
From the work of Lee and Yang\cite{LY2}, one finds that
$\xi=\frac{12}{5}(a/\lambda)(g_{3/2}^{2}(1)/g_{5/2}(1))$ 
for the  homogenous dilute Bose gas, which is around 
0.3 for the temperature range studied\cite{Griffin}.
{\bf (5)} In terms of dimensionless quantities 
$[\delta \tilde{P}, \delta\tilde{T}] \equiv 
[\delta P/(n_{o}({\bf r})k_{B}T_{o}), \delta T/T_{o}]$, we find that 
near $T_{c}$, $\delta\tilde{T}/\delta\tilde{P}\gg 1$ for all 
second sound modes, whereas $\delta\tilde{P}\sim 
\delta\tilde{T}$ for the first sound modes. 

{\bf Comparison with the JILA data}: Examining the JILA data\cite{Jin} on the 
sound modes of $^{87}$Rb with $\sim 2\times 10^3$ atoms, we
find surprising consistencies with the behaviors of the 
larger systems that we studied :
{\bf (a)} An $(m=0)$ mode with frequency $\approx 2\omega_{T}$ was 
observed for all $T$ above $T_{c}$\cite{Jin}. 
The analog of this mode in a spherical 
trap is the $(\ell=0, n_{1}=1)$ first sound mode, which also has 
frequency $\approx 2\omega_{T}$ for all $T$ above $T_{c}$. 
{\bf (b)} Below $T_{c}$, the frequency of the $(m=0)$ mode 
falls from about $2\omega_{T}$ to $1.85\omega_{T}$ as $T$ decreases from 
$0.9T_{co}$ to $0.5T_{co}$\cite{Jin}.  
The first and second sound analogs of this mode below
$T_{c}$ are the $(\ell=0, n_{1}=1)$ and the $(\ell=0, n_{2}=1)$ modes
respectively. 
The observed behavior matches well with the $(\ell=0, n_{2}=1)$ 
second sound mode, which drops from about 
$1.9\omega_{T}$ to $1.5\omega_{T}$ as $T$ decreases
from $0.8T_{co}$ to $0.6T_{co}$ as seen from Fig.1. 
{\bf (c)} An $(m=2)$ mode was also observed below $T_{c}$\cite{Jin}. 
Its frequency decreases from $1.45\omega_{T}$ to $1.25\omega_{T}$ as 
$T$ increases from $0.5T_{co}$ to $0.85T_{co}$. The second sound analog of 
this mode in spherical trap is the $(\ell=2, n_{2}=0)$ mode, which 
also drops from about $1.4\omega_{T}$ to about $1.35\omega_{T}$ as 
$T$ increases from $0.5T_{co}$ to $0.85T_{co}$.

While we do not expect perfect agreement of our results with 
the JILA observations\cite{Jin} 
because of the difference in trap symmetry and particle numbers,
the qualitative and quantitative consistencies over the temperature and
angular momentum range mentioned above are striking.
The above discussions suggest that the $(m=0)$ and $(m=2)$ modes observed 
below $T_{c}$ \cite{Jin} are both second sound modes. It is not clear at
present why the first sound modes do not appear with great prominence below
$T_{c}$. Whether it is due to the way that the modes are excited or due to 
the fact that density oscillations below $T_{c}$ might contain a large second
sound component because of the large temperature fluctuations in the second
sound modes (as mentioned in Discussion {\bf (5)} above) will be studied later. 
To clearly identify the nature of the sound modes, 
it is necessary to experimentally investigate  larger number of 
modes so as to have more consistency checks with the hydrodynamic predictions. 
We hope that this work will stimulate and provide guidance for future 
experiments. 

VBS would like to thank Allan McLeod for providing the programs on $g_{\nu}$, 
and  Vijay Shenoy and Shiwei Zhang for discussions on numerical methods.
Various parts of this work were performed during TLH's regular visits to CalTech
during Winter and Spring of 97. He would specially like to thank M.C. Cross 
for hospitality.  This work is supported by NSF Grant No. DMR-9705295.

\newpage 

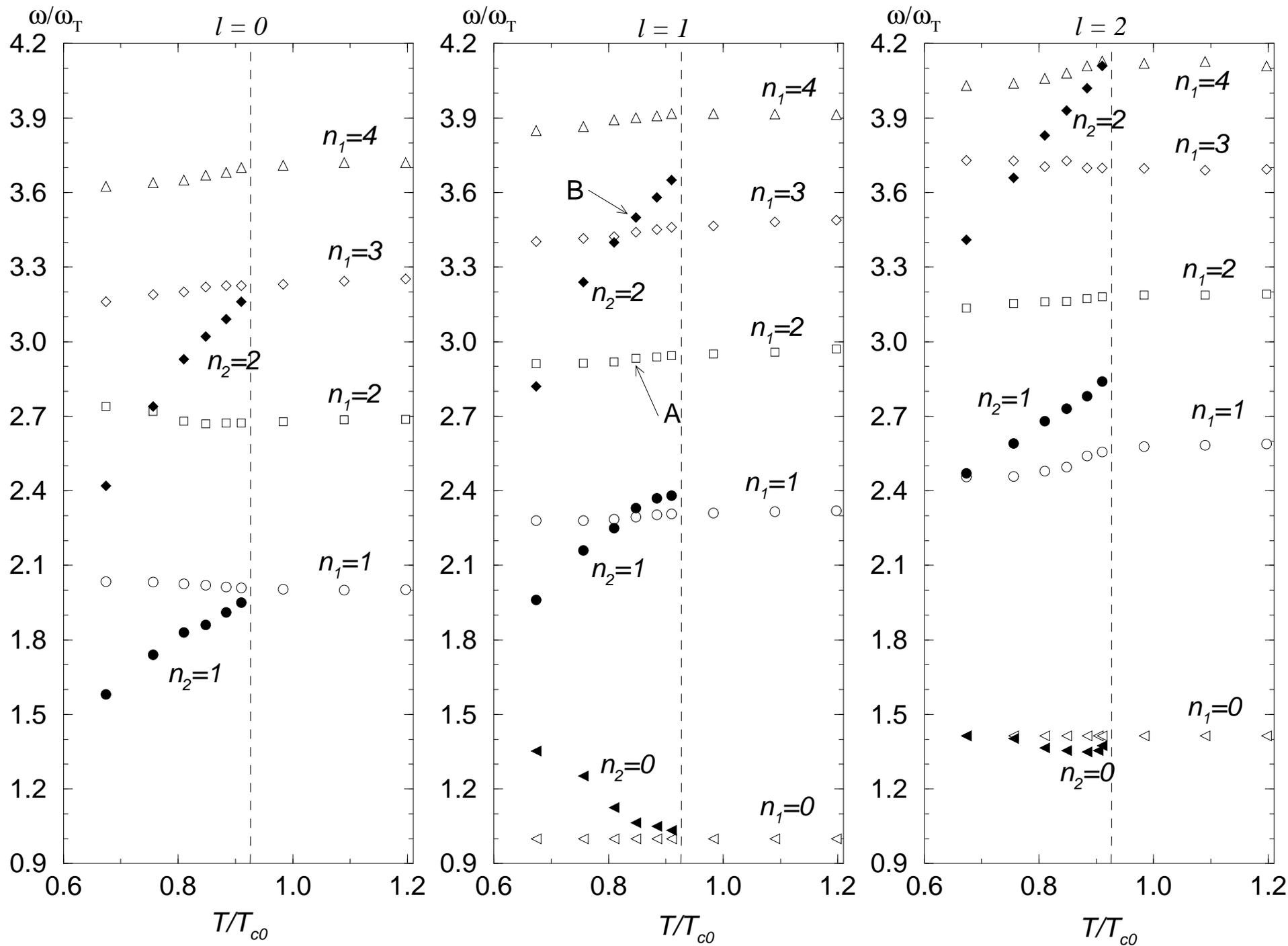
\noindent {\bf Figure 1:} The frequencies $\omega^{(1)}_{\ell, n_{1}}$ and 
$\omega^{(2)}_{\ell, n_{2}}$ are represented as 
open and filled symbols resp. They are plotted as a 
function of $T/T_{co}$, where $T_{c0}$ is the transition temperature of the
ideal Bose gas in the trap. The dotted line at $0.917 T_{co}$
indicates the critical temperature $T_c$ of the interacting model. 
For the $\ell\neq 0$ modes, $r=0$ is not counted as a node. 
As far as we can tell, $\omega_{\ell=0, n_{2}=0}=0$. 

\vspace{0.2in}

\noindent {\bf Figure 2a:} The eigenfunctions of $\omega^{(1)}_{\ell=1, 
n_{1}=2}$ (marked as $A$ in Fig.1) :
To magnify the features of the first sound, we have multiplied 
$\delta P$ and $\delta T$ by $r^2$ in Fig.2a. 

\vspace{0.2in}

\noindent {\bf Figure 2b:} The eigenfunctions of $\omega^{(2)}_{\ell=1,
n_{2}=2}$ (marked as $B$ in Fig.1) : 
These functions are not multiplied  by $r^2$ as those in Fig.2a because their 
features are sufficiently clear.  

\vspace{0.2in}

\noindent {\bf Fig.2c :}  The detailed structure in Fig.2b near 
$r^{\ast}$: The filled point indicates the location of $r^{\ast}$, and 
$\Delta$ is $0.005a_T$.  The sharp 
change of slope at $r^{\ast} \pm \Delta$ is expected as our boundary 
condition is meant to simulate the collapsing process of LDA at $r^{\ast}$. 

\vspace{0.2in}

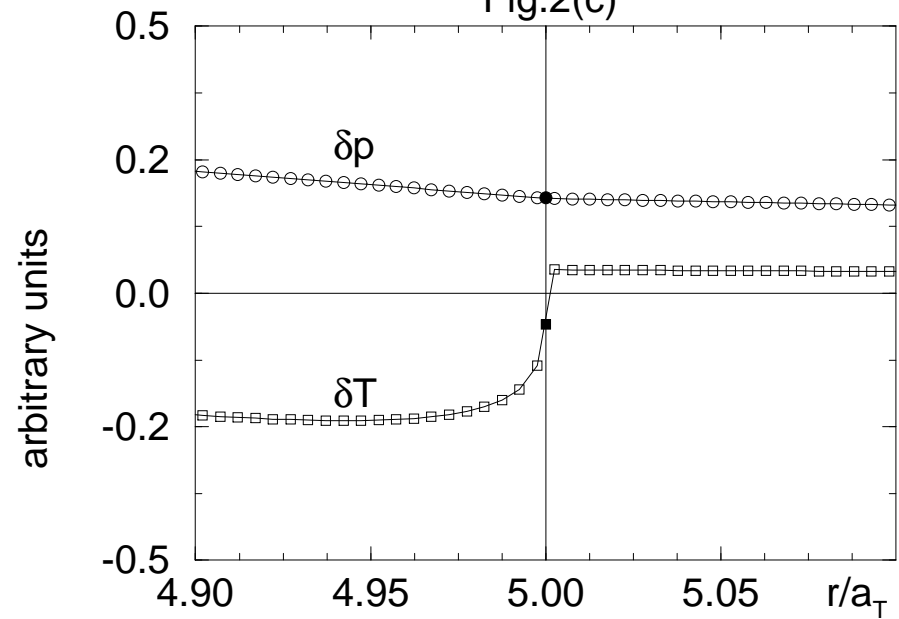
\noindent {\bf Figure 3:}  The $N$ dependence of $\omega^{(2)}_{\ell, n_{2}=1}$
modes $(\ell,n_{2}=1)$ near $T_c$. 
The temperature is chosen so that $r^{\ast}=a_{T}$.


\begin{references}
\bibitem{accurate} For a given grid spacing, all 
eigenstates except a set denoted as 
$\{ \omega^{(2)}_{\ell, n_{2}=0}\}$ below are accurate upto
one part in $10^4$, while those of the set are accurate upto one part in $10^3$.
\end{references}
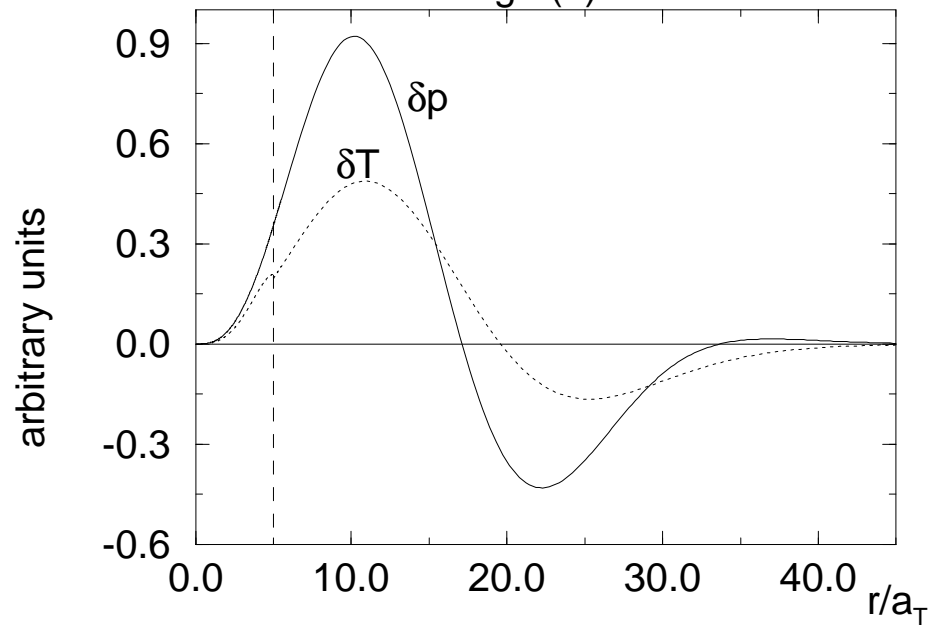
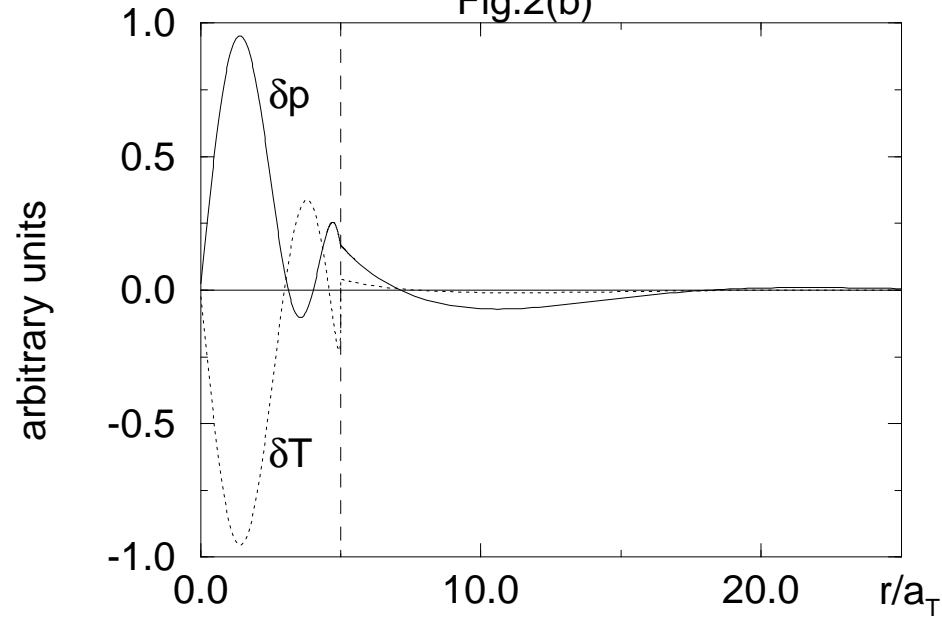
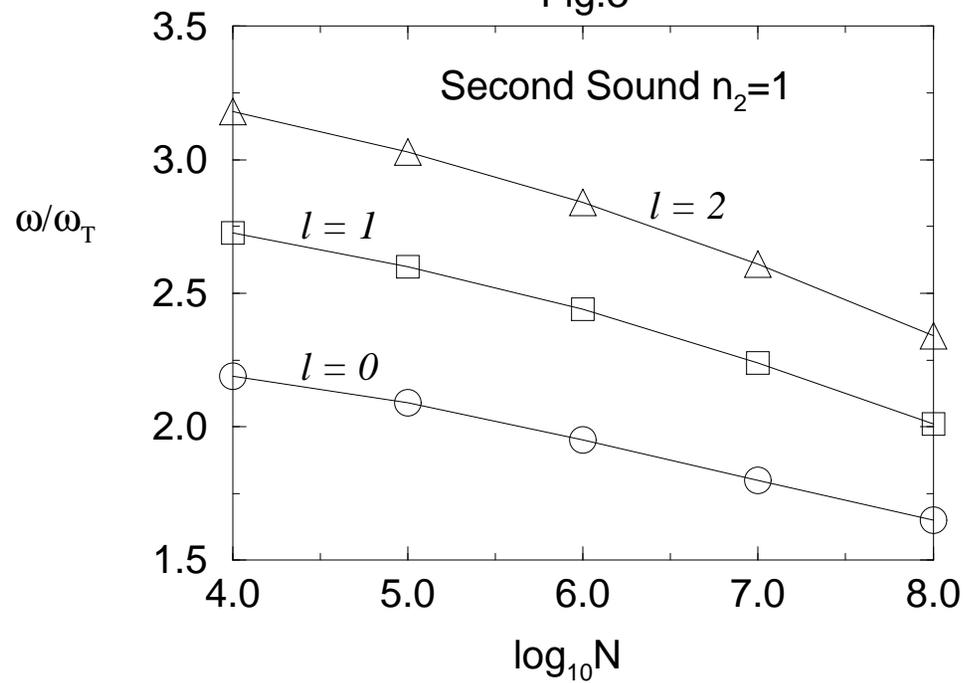
\end{document}